\documentclass[runningheads]{llncs}
\usepackage[T1]{fontenc}
\usepackage{graphicx}
\usepackage[T1]{fontenc}
\usepackage{graphicx}
\usepackage{amsmath, amssymb}
\usepackage{algorithm}
\usepackage{algpseudocode}
\usepackage{multirow}
\usepackage{booktabs}
\usepackage{lipsum}
\usepackage{array}
\usepackage{listings}
\usepackage[colorlinks=true,linkcolor=blue,citecolor=blue,urlcolor=blue,]{hyperref}
\usepackage[table]{xcolor}
\usepackage{marvosym}

\begin{document}
\title{Striving for Simplicity: Simple Yet Effective Prior-Aware Pseudo-Labeling for Semi-Supervised Ultrasound Image Segmentation}
%
%
\author{Yaxiong Chen\inst{1,2} \and 
Yujie Wang\inst{1}\thanks{Work done during an internship at MedAI Technology (Wuxi) Co. Ltd.} \and
Zixuan Zheng\inst{3} \and
Jingliang Hu\inst{3} \and
Yilei Shi\inst{3} \and
Shengwu Xiong\inst{1,2} \and
Xiao Xiang Zhu\inst{4} \and
Lichao Mou\inst{3}\textsuperscript{(\Letter)}}


%
\authorrunning{Y. Chen et al.}
%
\institute{Wuhan University of Technology, Wuhan, China \and Shanghai Artificial Intelligence Laboratory, Shanghai, China \and MedAI Technology (Wuxi) Co. Ltd., Wuxi, China\\\email{lichao.mou@medimagingai.com} \and Technical University of Munich, Munich, Germany}

\maketitle              
\begin{abstract}
Medical ultrasound imaging is ubiquitous, but manual analysis struggles to keep pace. Automated segmentation can help but requires large labeled datasets, which are scarce. Semi-supervised learning leveraging both unlabeled and limited labeled data is a promising approach. State-of-the-art methods use consistency regularization or pseudo-labeling but grow increasingly complex. Without sufficient labels, these models often latch onto artifacts or allow anatomically implausible segmentations. In this paper, we present a simple yet effective pseudo-labeling method with an adversarially learned shape prior to regularize segmentations. Specifically, we devise an encoder-twin-decoder network where the shape prior acts as an implicit shape model, penalizing anatomically implausible but not ground-truth-deviating predictions. Without bells and whistles, our simple approach achieves state-of-the-art performance on two benchmarks under different partition protocols. We provide a strong baseline for future semi-supervised medical image segmentation. Code is available at \url{https://github.com/WUTCM-Lab/Shape-Prior-Semi-Seg}.

\keywords{semi-supervised learning  \and segmentation \and pseudo-labeling.}
\end{abstract}
\section{Introduction}
Medical ultrasound imaging is one of the most ubiquitous diagnostic modalities used in clinical practice today owing to its non-invasive, radiation-free, and real-time nature~\cite{xie2022uncertainty,ning2022smunet}. This ubiquity has resulted in an explosion of ultrasound images. However, manual screening by clinical experts struggles to keep pace. Automated ultrasound image segmentation models can improve efficiency and consistency, but require large pixel-wise labeled datasets which are particularly scarce in the ultrasound domain compared to natural images or even other medical imaging modalities like CT and MRI~\cite{chen2023deep}.
\par

\begin{figure}[!th]
\centering
\includegraphics[width=\textwidth]{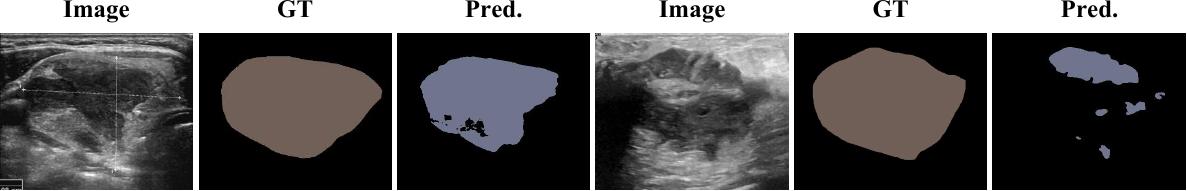}
\caption{Examples of unrealistic and unnatural segmentations produced by a UNet.} \label{fig:motivation}
\end{figure}

A common modus operandi is to use semi-supervised learning, that is, leveraging both abundant unlabeled medical images and limited labeled ones simultaneously~\cite{xu2022shadow}. The majority of current state-of-the-art methods follow two schemes: consistency regularization and pseudo-labelling~\cite{xu2022bayesian}. The consistency regularization-based approaches incorporate a regularization term in the loss to minimize the discrepancy between different predictions for the same image, which are derived by introducing perturbations to the input image or to the models involved. \cite{bortsova2019semi} proposes a network with two identical branches, each of which receives the same image with different perturbations, and designs a loss to encourage predictions of the two branches to be consistent. Subsequently, the authors of~\cite{ouali2020semi} introduce more branches, devise distinct perturbations for each of them, and enforce consistency regularization across branches. Further, \cite{zhong2023semi} presents a sophisticated attention mechanism for the multi-branch architecture. On the other hand, there has been a significant amount of recent work based on pseudo-labelling. For instance,~\cite{tarvainen2017mean} proposes a method called mean teacher, which adopts a teacher-student architecture for semi-supervised learning. This framework comprises a teacher model and a student model. The student model is trained on unlabeled data using pseudo-labels generated by the teacher, and the teacher model itself is an exponential moving average of the student model's weights. In~\cite{zhang2021robust}, the authors present mutual learning based on the mean teacher framework and introduce an additional self-correction mechanism to refine pseudo-labels. \cite{kwon2022semi} further devises an error localization module to help identify incorrect pixels in pseudo labels. Nevertheless, albeit successful, these state-of-the-art methods tend to increasingly complex designs as diverse as more network modules and additional losses.
\par

Moreover, we observe that without sufficient labeled ultrasound images, models often latch onto artifacts rather than meaningful morphological cues. In addition, the flexibility of deep networks to fit arbitrary boundaries allows unrealistic and unnatural segmentations that violate basic anatomical principles (cf. Fig.~\ref{fig:motivation}). These pose challenges for semi-supervised image segmentation approaches. For example, consistency regularization-based methods encourage predictions to be invariant to perturbations, but are unable to correct errors due to anatomically implausible boundaries in the absence of ground truth constraints. The pseudo-labeling methodology relies on predictions from the teacher model to generate targets for the student, which may reinforce erroneous cues in areas of high uncertainty. To alleviate the issues, one possible solution is to utilize shape information. \cite{oktay2017anatomically} and \cite{10.1007/978-3-030-32226-7_65} are motivated to improve segmentation masks generated by fully supervised networks by better matching them to ground truths, but they do not take advantage of anatomical shape prior knowledge, which is crucial for semi-supervised ultrasound image segmentation where boundaries of targets are often ambiguous.
\par

\begin{figure}[!t]
    \centering
    \includegraphics[width=0.85\textwidth]{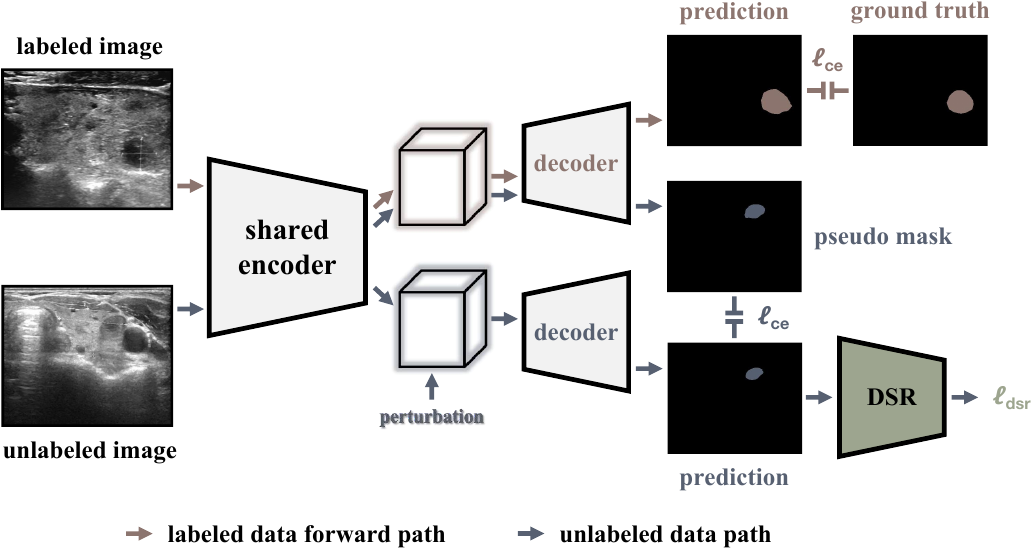}
    \caption{Pipeline of the proposed semi-supervised method for ultrasound image segmentation.}
    \label{fig:method}
\end{figure}

In this paper, we break the trend of recent state-of-the-art approaches that combine increasingly complex techniques, and introduce a simple and clean pseudo-labeling-based network. To address the problems caused by artifacts and ambiguous boundaries of lesions in ultrasound images, we learn a shape prior and inject this prior into the network. Specifically, our contributions are three-fold.
\begin{itemize}
    \item Without complex designs, a simple yet effective encoder-twin-decoder network architecture is devised, which can readily achieve better performance.
    \item By integrating an adversarially learned shape prior, we push the limit of semi-supervised ultrasound image segmentation. The learned prior serves as a regularizer, penalizing the network only if its outputs are unrealistic, not if it deviates from ground truths.
    \item We provide a simple yet strong baseline for future works. Extensive experiments and ablation studies are conducted on two public ultrasound datasets to demonstrate its effectiveness.
\end{itemize}

\section{Method}
We first present the overall structure of our proposed network in Section~\ref{PSTNet}. In Section~\ref{spm}, we then introduce shape prior modeling and the shape regularizer that is incorporated into the network. Finally, the details of the training process for the entire network are described in Section~\ref{TS}.

\subsection{Semi-Supervised Image Segmentation Network}
\label{PSTNet}
We propose a semi-supervised ultrasound image segmentation network that concurrently leverages labeled and unlabeled data for training. As illustrated in Fig.~\ref{fig:method}, our network is composed of three key components---an encoder $\mathcal{E}$, twin decoders $\mathcal{D}_{l}$ and $\mathcal{D}_{p}$, and a deep shape regularizer (DSR). $\mathcal{E}$ is a ResNet~\cite{He_2016_CVPR} that maps an input image into feature representations. The twin decoders, $\mathcal{D}_l$ and $\mathcal{D}_p$, share an identical architecture with five convolutional blocks, each containing two Conv-BN-ReLU layers, but serve distinct purposes. Specifically, $\mathcal{D}_l$ focuses exclusively on learning from limited labeled data to produce pseudo-labels for supervising $\mathcal{D}_p$. In contrast, $\mathcal{D}_p$ functions in a prior-guided manner to segment unlabeled images by integrating guidance from the pseudo-labels and regularization from DSR, without accessing any labeled data during training. Notably, only $\mathcal{E}$ and $\mathcal{D}_p$ are required at inference time. In essence, the proposed method seamlessly integrates semi-supervised learning and inductive bias modeling for segmentation in an end-to-end framework.

\subsection{Shape Prior Modeling and Shape Regularization}
\label{spm}
\subsubsection{Shape Prior Modeling}
We pre-train a generative adversarial network (GAN) to determine whether a shape is anatomically plausible. The GAN contains a generator that synthesizes shape masks from a randomly sampled latent vector, and a discriminator that distinguishes the generated masks from real ones. Through this adversarial training process, the discriminator learns to model the distribution of real shapes and has the ability to quantify the plausibility of any input mask. We therefore utilize the trained discriminator as a regularizer, namely DSR, for subsequent semi-supervised image segmentation. Note that the generator is discarded after pre-training.
\par
Specifically, we use five transposed convolutional layers with batch normalization (BN) and ReLU for the generator, and five convolutional layers with leaky ReLU for the discriminator. Due to the limited number of ground truth segmentation masks, stably training the GAN is challenging. Additionally, the high resolution of the masks leads to training instability. To mitigate this, we uniformly resize all masks to $64\times64$, which does not alter the distribution of shapes yet eases optimization. Further, following~\cite{gulrajani2017improved}, we use the Wasserstein loss with a gradient penalty term to train our GAN. The loss is defined as:
\begin{equation}\label{wganloss}
\mathcal{L}_{\mathrm{spm}}=\underset{\tilde{x} \sim P_{g}}{\mathbb{E}}[D(\tilde{x})]-\underset{x \sim P_{r}}{\mathbb{E}}[D(x)]+\lambda \underset{\hat{x} \sim P_{x}}{\mathbb{E}}[\left(\left\|\nabla_{\hat{x}} D(\hat{x})\right\|_{2}-1\right)^{2}] \,,
\end{equation}
where $D$ is the discriminator, and $P_g$ and $P_r$ represent distributions of the generated and real masks, respectively. The last term refers to the gradient penalty introduced in~\cite{gulrajani2017improved}.

\subsubsection{DSR}
At training time, the discriminator aims to minimize $\mathcal{L}_{\mathrm{spm}}$, while the generator attempts to maximize $\mathcal{L}_{\mathrm{spm}}$ by fooling the discriminator via the first term in Eq.~(\ref{wganloss}). After training, the discriminator gains the ability to assess the plausibility of shape masks. Therefore, we can define DSR as
\begin{equation}\label{dsr}
\ell_{\mathrm{dsr}}=-\underset{\tilde{x} \sim P_{m}}{\mathbb{E}}[D(\tilde{x})] \,,
\end{equation}
where $P_{m}$ denotes the distribution of segmentation masks predicted by the semi-supervised image segmentation network.

\subsection{Training Strategy and Loss
}\label{TS}
We detail the training process of the proposed semi-supervised image segmentation network. In each iteration, we sample a batch of labeled images $\mathcal{B}_{s}=\left\{\left(x_{i}, y_{i}\right)\right\}_{i=1}^{\left|\mathcal{B}_{s}\right|}$ and a batch of unlabeled images $\mathcal{B}_{u}=\left\{\left(u_{i}\right)\right\}_{i=1}^{\left|\mathcal{B}_{u}\right|}$.
\par
First, we pass each labeled image $x_i\in\mathcal{B}_s$ through the encoder $\mathcal{E}$ to attain feature maps, which are then fed into the decoder $\mathcal{D}_{l}$ to generate a segmentation mask. A cross entropy loss is used to measure the discrepancy between this mask and the corresponding ground truth:
\begin{equation}\label{loss1}
\mathcal{L}_s=\frac{1}{|\mathcal{B}_s|}\sum^{|\mathcal{B}_s|}_{i=1}\ell_{\mathrm{ce}}(\mathcal{D}_l\circ\mathcal{E}(x_i),y_i) \,,
\end{equation}
where $\circ$ is a composition function.
\par
Next, for each unlabeled image $u_i\in\mathcal{B}_{u}$, the encoder $\mathcal{E}$ extracts feature representations, and $\mathcal{D}_{l}$ is used to produce a pseudo segmentation mask. Afterwards, we apply dropout on the feature representations and pass the perturbed features through $\mathcal{D}_p$ to predict a segmentation mask. A consistency loss is exploited to enforce agreement between the two masks. In addition, DSR evaluates the plausibility of the predicted segmentation map and outputs a score that imposes a constraint on learning $\mathcal{D}_p$:
\begin{equation}\label{loss2}
\mathcal{L}_u=\frac{1}{|\mathcal{B}_u|}\sum^{|\mathcal{B}_u|}_{i=1}\ell_{\mathrm{ce}}(\mathcal{D}_p\circ m_{i}\odot\mathcal{E}(u_i),\mathcal{D}_l\circ\mathcal{E}(u_i))+\lambda\ell_{\mathrm{dsr}}(\mathcal{D}_p\circ m_{i}\odot\mathcal{E}(u_i)) \,,
\end{equation}
where $\lambda$ controls the strength of the regularization, and $m_i$ is the dropout's mask.
\par
Finally, the loss of our method is defined as
\begin{equation}\label{loss}
\mathcal{L}=\mathcal{L}_s+\gamma\mathcal{L}_u \,.
\end{equation}

\section{Experiments}
\subsection{Datasets and Evaluation Metrics}
To validate the effectiveness and robustness of our network, we conduct experiments on two ultrasound image segmentation datasets. The TN3K dataset~\cite{GONG2023106389} contains 3,493 thyroid nodule ultrasound images, of which we use 614 for testing, 301 for validation, and the remaining 2,578 for training. The BUSI dataset~\cite{ALDHABYANI2020104863} collects 780 breast ultrasound images with an average size of 500×500 pixels. The images are categorized into normal, benign, and malignant classes. We use the 647 benign and malignant images, with 576 in the training set and 71 in the test set.
\par
Following common partition protocols in semi-supervised medical image segmentation, we perform experiments using $1/8$, $1/4$, and $1/2$ of the total labeled data for each dataset. Moreover, we adopt two metrics, Dice similarity coefficient and intersection over union (IoU), to quantify segmentation performance.

\begin{table*}[!t]
\centering
\caption{Comparison with state-of-the-art methods on the TN3K dataset under different partition protocols. The results of the fully supervised baseline are marked in \colorbox{gray!20}{gray}.}
\begin{center}
\renewcommand\arraystretch{1.0}
\setlength{\tabcolsep}{4pt}
\begin{tabular}{l cccccc}
 & \multicolumn{2}{c}{1/8} & \multicolumn{2}{c}{1/4} &\multicolumn{2}{c}{1/2} \\
\midrule[0.85pt]
Methods & Dice & IoU & Dice & IoU & Dice & IoU \\
\midrule[0.5pt]
\rowcolor{gray! 20} 
UNet~\cite{10.1007/978-3-319-24574-4_28} & 73.98 & 62.95 & 80.06 & 70.61 & 81.75 & 72.89 \\
FixMatch~\cite{NEURIPS2020_06964dce} & 61.59 &49.36 & 66.09 & 53.71 & 69.44 & 57.74 \\
SemiMedSeg~\cite{10.1007/978-3-030-87196-3_13} & 77.33 & 67.11 & 80.67 & 71.44 & 81.79 & 72.71 \\
U2PL~\cite{Wang_2022_CVPR} & 74.35 & 63.15 & 75.60 & 64.26 & 75.25 & 63.89 \\
ST++~\cite{Yang_2022_CVPR} & 52.51 & 41.12 & 57.61 & 46.05 & 65.21 & 53.54 \\
AugSeg~\cite{Zhao_2023_CVPR} & 72.82 & 61.38 & 74.10 & 63.22 & 76.09 & 65.28 \\
UniMatch~\cite{Yang_2023_CVPR} & 67.30 & 55.25 & 69.33 & 57.29 & 69.67 & 57.88 \\
MTNet~\cite{zhong2023semi} & 61.26 & 51.33 & 68.77 & 58.45 & 73.72 & 63.62 \\
SS-Net~\cite{10.1007/978-3-031-16443-9_4} & 70.99 & 59.90 & 72.93 & 62.80 & 73.49 & 63.89 \\
Ours & \textbf{82.21} & \textbf{73.16} & \textbf{83.24} & \textbf{74.47} & \textbf{84.17} & \textbf{75.70} \\
\bottomrule[0.85pt]
\end{tabular}
\end{center}
\label{tab:tab1}
\end{table*}

\begin{table*}[!t]
\centering
\caption{Comparison with state-of-the-art methods on the BUSI dataset.}
\begin{center}
\renewcommand\arraystretch{1.0}
\setlength{\tabcolsep}{4pt}
\begin{tabular}{l cccccc}
 & \multicolumn{2}{c}{1/8} & \multicolumn{2}{c}{1/4} &\multicolumn{2}{c}{1/2} \\
\midrule[0.85pt]
Methods & Dice & IoU & Dice & IoU & Dice & IoU \\
\midrule[0.5pt]
\rowcolor{gray! 20} 
UNet~\cite{10.1007/978-3-319-24574-4_28} & 67.89 & 57.68 & 77.01 & 67.60 & 77.06 & 70.88 \\
FixMatch~\cite{NEURIPS2020_06964dce} & 42.76 & 34.96 & 50.35 & 41.10 & 56.44 & 46.87 \\
SemiMedSeg~\cite{10.1007/978-3-030-87196-3_13} & 72.04 & 63.27 & 77.73 & 69.05 & 77.03 & 68.71 \\
U2PL~\cite{Wang_2022_CVPR} & 68.50 & 59.03 & 68.97 & 60.55 & 72.76 & 64.62 \\
ST++~\cite{Yang_2022_CVPR} & 52.81 & 40.78 & 61.67 & 50.10 & 63.09 & 52.28 \\
AugSeg~\cite{Zhao_2023_CVPR} & 64.78 & 54.76 & 68.54 & 59.24 & 72.03 & 63.64 \\
UniMatch~\cite{Yang_2023_CVPR} & 53.96 & 43.69 & 55.87 & 45.53 & 57.03 & 46.52 \\
MTNet~\cite{zhong2023semi} & 51.30 & 41.10 & 60.57 & 50.47 & 68.89 & 58.60 \\
SS-Net~\cite{10.1007/978-3-031-16443-9_4} & 61.71 & 53.30 & 66.90 & 57.32 & 70.61 & 62.26 \\
Ours & \textbf{75.19} & \textbf{66.90} & \textbf{79.43} & \textbf{71.17} & \textbf{80.41} & \textbf{72.02} \\
\bottomrule[0.85pt]
\end{tabular}
\end{center}
\label{tab:tab2}
\end{table*}

\subsection{Implementation Details}
\subsubsection{DSR Training} We first train a GAN leveraging existing labels. Specifically, we resize all ground truth segmentation masks into a fixed size of $64\times 64$ and set batch size to 16. We optimize the generator and discriminator using two RMSprop optimizers with a learning rate of 0.00005, and the total number of epochs is set to 5,000.

\begin{figure}[!t]
\centering
\includegraphics[width=\textwidth]{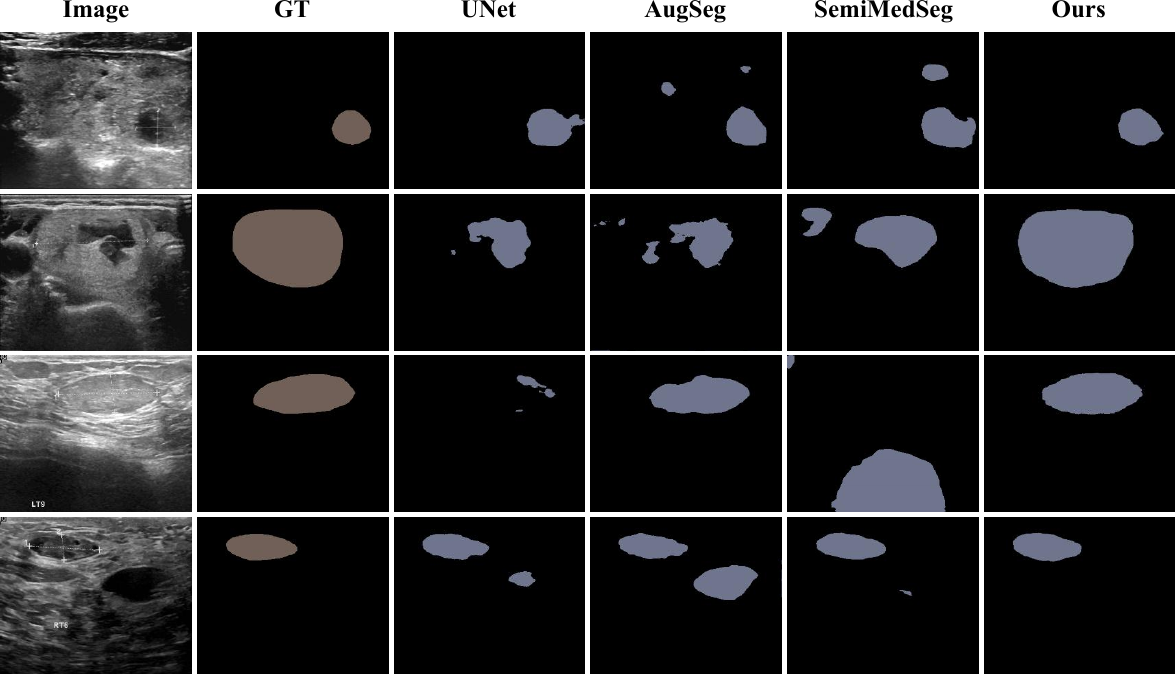}
\caption{Qualitative comparison of segmentation results produced by our method and other competitors on the TN3K and BUSI test sets. The top two rows show examples from TN3K, while the bottom two rows show cases from BUSI.} \label{fig:view}
\end{figure}

\subsubsection{Segmentation Network Training}
After training DSR, we train the segmentation network by resizing all images to $320\times320$ pixels, followed by random cropping to $256\times256$ pixels and applying random data augmentation (rotation/scaling). We use a batch size of 16 and SGD with a momentum of 0.9 and a weight decay of 0.00001 to optimize the model. A polynomial learning rate schedule is adopted to adjust the learning rate $lr= init\_lr \times \left(1-\frac{iter}{max\_iter}\right)^{power}$, where $init\_lr = 0.001$ and $power = 0.9$. The total number of epochs is set to 200.

\subsection{Comparison with State-of-the-Art Methods}
In our experiments, we use a fully supervsed UNet~\cite{10.1007/978-3-319-24574-4_28} (with ResNet34 backbone) as the baseline. We compare our approach with current state-of-the-art semi-supervised segmentation models including FixMatch~\cite{NEURIPS2020_06964dce}, SemiMedSeg~\cite{10.1007/978-3-030-87196-3_13}, U2PL~\cite{Wang_2022_CVPR}, ST++~\cite{Yang_2022_CVPR}, AugSeg~\cite{Zhao_2023_CVPR}, UniMatch~\cite{Yang_2023_CVPR}, MTNet~\cite{zhong2023semi}, and SS-Net~\cite{10.1007/978-3-031-16443-9_4}. 

Table~\ref{tab:tab1} and Table~\ref{tab:tab2} show experimental results on the TN3K and BUSI datasets. Our approach achieves the best performance, especially under sparse labeling (1/8). We improve 8.23\% and 4.88\% over the fully supervised baseline and second best semi-supervised image segmentation method on TN3K. Similar significant gains are observed on BUSI with limited labels.

Fig.~\ref{fig:view} shows some visual examples. We can see that our approach produces high-quality segmentations on large objects, small objects, and with multiple instances, outperforming other methods which may occasionally produce segmentation errors.

\begin{table*}[!t]
\centering
\caption{Ablation study quantifying the contribution of DSR on the TN3K and BUSI datasets under three data partition protocols.}
\begin{center}
\renewcommand\arraystretch{1.0}
\setlength{\tabcolsep}{4pt}
\begin{tabular}{llcccccc}
 && \multicolumn{2}{c}{1/8} & \multicolumn{2}{c}{1/4} &\multicolumn{2}{c}{1/2} \\
\midrule[0.85pt]
Datasets & Methods & Dice & IoU & Dice & IoU & Dice & IoU \\
\midrule[0.5pt]
\multirow{2}{*}{TN3K} & w/o DSR & 81.19 & 72.17 & 82.26 & 73.34 & 83.08 & 74.63 \\
& Full model &\textbf{82.21} & \textbf{73.16} & \textbf{83.24} & \textbf{74.47} & \textbf{84.17} & \textbf{75.70} \\
\midrule[0.5pt]
\multirow{2}{*}{BUSI}& w/o DSR & 74.14 & 65.89 & 78.31 & 69.23 & 79.51 & 71.13 \\
&Full model & \textbf{75.19} & \textbf{66.90} & \textbf{79.43} & \textbf{71.17} & \textbf{80.41} & \textbf{72.02} \\
\bottomrule[0.85pt]
\end{tabular}
\end{center}
\label{tab:tab3}
\end{table*}

\begin{figure}[!t]
\centering
\includegraphics[width=0.85\textwidth]{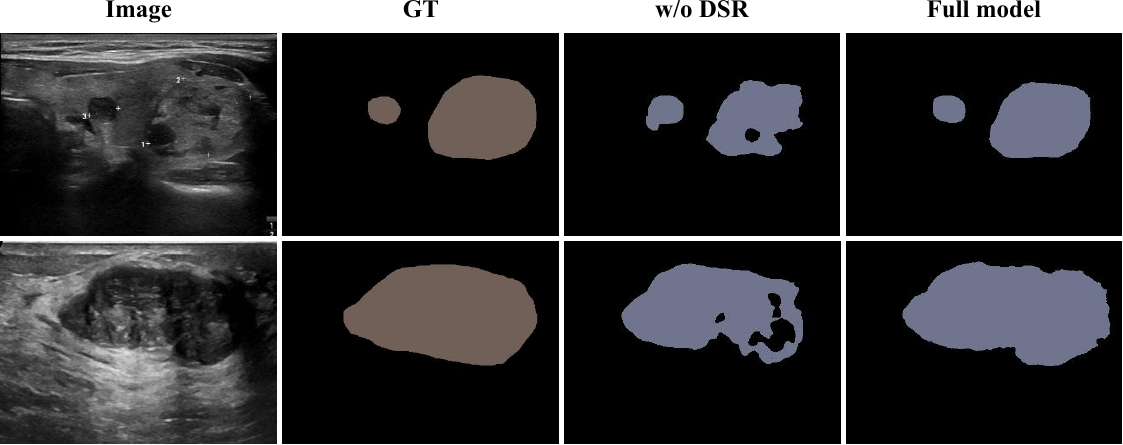}
\caption{Qualitative assessment of our proposed DSR module on the TN3K and BUSI test sets. The top row shows an example from TN3K, while the bottom case is from BUSI.} \label{fig:view2}
\end{figure}

\subsection{Ablation Study}
To validate the effectiveness of DSR, we conduct ablation studies on TN3K and BUSI. As summarized in Table~\ref{tab:tab3}, incorporating DSR consistently improves performance by approximately 1\% on both datasets. This indicates that exploiting shape priors helps alleviate issues like porous. Qualitative results in Fig.~\ref{fig:view2} showcase segmentation masks with more natural, coherent shapes when using DSR.

\section{Conclusion}
In this work, we present a simple yet effective pseudo-labeling framework with an adversarially learned shape prior for semi-supervised medical ultrasound image segmentation. Unlike recent studies that tend to combine increasingly complex mechanisms, our approach utilizes a clean encoder-twin-decoder network architecture. We demonstrate the model on two public benchmarks and qualitatively and quantitatively establish the proficiency of the framework. The experimental results show that, without requiring complex architectures or augmentation strategies, our method achieves state-of-the-art performance under different data partition protocols.

\begin{credits}
\subsubsection{\ackname} This work is supported in part by the National Key Research and Development Program of China (2022ZD0160604), in part by the Natural Science Foundation of China (62101393/62176194), in part by the High-Performance Computing Platform of YZBSTCACC, and in part by MindSpore (\url{https://www.mindspore.cn}), a new deep learning framework.

\subsubsection{\discintname}
The authors have no competing interests to declare that are relevant to the content of this paper.
\end{credits}
%
%
%
%

\end{document}